\documentclass[conference]{IEEEtran}
\IEEEoverridecommandlockouts

\usepackage{amsmath,amsfonts}
\usepackage{algorithmic}
\usepackage{algorithm}
\usepackage{array}
\usepackage[caption=false,font=normalsize,labelfont=sf,textfont=sf]{subfig}
\usepackage{textcomp}
\usepackage{stfloats}
\usepackage{url}
\usepackage{verbatim}
\usepackage{graphicx}
\usepackage{cite}
\usepackage{color}

\usepackage{amssymb}
\newtheorem{definition}{Definition}

\newtheorem{proposition}[definition]{Proposition}
\newtheorem{lemma}[definition]{Lemma}
\newtheorem{theorem}[definition]{Theorem}
\newtheorem{corollary}[definition]{Corollary}
\newtheorem{example}[definition]{Example}
\newtheorem{remark}[definition]{Remark}
\newcommand {\beq}[1]{\begin{equation}
                      \label{eq:#1} }
\newcommand {\eeq}{\end{equation}}

\newcommand {\req}[1]{(\ref{eq:#1})}

\newcommand {\bear}[1]{
                       \begin{eqnarray}
                       \label{eq:#1} }
\newcommand {\eear}{\end{eqnarray}}

\newcommand {\bearn}{\begin{eqnarray*}}
\newcommand {\eearn}{\end{eqnarray*}}
\newcommand {\bsec}[2]{
                       \section{#1}
                       \label{sec:#2} }

\newcommand {\rsec}[1]{Section \ref{sec:#1}}
\newcommand {\bsubsec}[2]{
                       \subsection{#1}
                       \label{sec:#2} }

\newcommand {\rfig}[1]{Figure \ref{fig:#1}}

\newcommand {\bdefin}[1]{\begin{definition}
                             \label{def:#1} }
\newcommand {\edefin}       {\end{definition}}

\newcommand {\bprop}[1]{\begin{proposition}
                              \label{prop:#1} }
\newcommand {\eprop}       {\end{proposition}}

\newcommand {\blem}[1]{\begin{lemma}
                            \label{lem:#1} }
\newcommand {\elem}   {\end{lemma}}

\newcommand {\bthe}[1]{\begin{theorem}
                         \label{the:#1} }
\newcommand {\ethe}   {\end{theorem}}
\newcommand {\rthe}[1]{Theorem \ref{the:#1}}

\newcommand {\bproof}{\noindent {\bf Proof.} \ }
\newcommand {\eproof} {\hfill $\square$ \\ \vspace{.3cm}}
\newcommand {\bcor}[1]{\begin{corollary}
                           \label{cor:#1} }
\newcommand {\ecor}   {\end{corollary}}

\newcommand {\bex}[2]{\vspace{.1in}
                      \begin{example}
                                {\bf #2}
                      \label{ex:#1} }
\newcommand {\eex}       {\end{example} \vspace{.3cm} }

\newcommand {\brem}[1]{\begin{remark}
                       \label{rem:#1} \em }
\newcommand {\erem}   {\end{remark}}

\def\ex{{\bf\sf E}}
\def\pr{{\bf\sf P}}

\begin{document}

\title{Fast Multichannel Topology Discovery in Cognitive Radio Networks}

\author{Yung-Li Wang, Yiwei Liu and Cheng-Shang Chang\\
Institute of Communications Engineering\\
National Tsing Hua University \\
Hsinchu 300044, Taiwan, R.O.C. \\
Email:  student0805@gmail.com; s112064521@m112.nthu.edu.tw; cschang@ee.nthu.edu.tw}

% The paper headers
%\markboth{Journal of \LaTeX\ IEEE COMMUNICATIONS LETTERS, ,~Vol.~xx, No.~x, June~2024}%
%{Shell \MakeLowercase{\textit{et al.}}: A Sample Article Using IEEEtran.cls for IEEE Journals}

%\IEEEpubid{0000--0000/00\$00.00~\copyright~2021 IEEE}
% Remember, if you use this you must call \IEEEpubidadjcol in the second
% column for its text to clear the IEEEpubid mark.

\maketitle

\begin{abstract}
In Cognitive Radio Networks (CRNs), secondary users (SUs) must efficiently discover each other across multiple communication channels while avoiding interference from primary users (PUs). Traditional multichannel rendezvous algorithms primarily focus on enabling pairs of SUs to find common channels without explicitly considering the underlying network topology. In this paper, we extend the rendezvous framework to explicitly incorporate network topology, introducing the \emph{multichannel topology discovery problem}. We propose a novel \emph{pseudo-random sweep algorithm with forward replacement}, designed to minimize correlation between consecutive unsuccessful rendezvous attempts, thereby significantly reducing the expected time-to-discovery (ETTD). Additionally, we introduce a \emph{threshold-based stick-together strategy} that dynamically synchronizes user hopping sequences based on partially known information, further enhancing discovery efficiency.
Extensive simulation results validate our theoretical analysis, demonstrating that the proposed algorithms substantially outperform conventional (sequential) sweep methods.
\end{abstract}


%{\bf \textit{Index Terms---}Multichannel rendezvous, locality-sensitive hashing, consistent functions.}
%\begin{IEEEkeywords}
%Multichannel rendezvous, locality-sensitive hashing, consistent hashing.
%\end{IEEEkeywords}

\section{Introduction}
\label{sec:introudction}

In the Cognitive Radio Networks (CRNs), wireless devices operate across multiple communication channels and must discover each other to initiate communication. Since primary users (PUs) may occupy certain frequency bands, secondary users (SUs) must identify spectrum holes to find channels available for communication. This fundamental challenge, known as the Multichannel Rendezvous Problem (MRP), involves two secondary users attempting to find a common channel by hopping across their available channels over time. As discussed extensively in prior studies, such as \cite{Theis2011,Bian2013,Book}, efficient solutions to the MRP are critical for neighbor discovery in many CRNs.

Most of the channel hopping (CH) sequences in the literature are targeted for minimizing the maximum time-to-rendezvous (MTTR) with maximum diversity, including CRSEQ \cite{CRSEQ}, JS \cite{JS2011}, DRDS \cite{DRDS13}, T-CH \cite{Matrix2015}, DSCR \cite{DSCR2016}, IDEAL-CH \cite{GAP2019}, and RDSML-CH \cite{Wang2022}. For a network with \(N\) channels, these CH sequences guarantee rendezvous within a period typically on the order of \(O(N^2)\). When the numbers of available channels of the two users are much smaller than $N$, it has been shown that the MTTR of the CH sequences in \cite{Chen14,Improved2015,Chang18} can be reduced to $O((\log\log N)n_1 n_2)$, where $n_1$ (resp. $n_2$) is the number of available channels of user 1 (resp. user 2).
However, the performance of the expected time-to-rendezvous (ETTR) of these CH sequences is not as good as that of the random algorithm (that randomly selects a channel from the available channel set). The ETTR of the random algorithm is  $n_1 n_2/n_{1,2}$, where $n_{1,2}$ is the number of common channels between the two users.

Recently,  Jiang and Chang \cite{LSH} proposed an innovative approach for the MRP by using the locality-sensitive hashing (LSH) technique. They utilized the practical observation that users are typically in close proximity and have similar available channel sets, resulting in a large Jaccard index between those sets.  The Jaccard index $J$ is  defined as $\frac{n_{1,2}}{n_{1} + n_{2} - n_{1,2}}$.
They showed that the ETTRs of their LSH algorithms can outperform the random algorithm when the Jaccard index between the two available channel sets is large. The LSH technique from \cite{LSH} was applied to CRNs without global channel enumeration in \cite{cheng2024multichannel}, and was later generalized to construct consistent CH sequences in \cite{CLC25}.

\begin{figure}[!t]
	\centering
	\includegraphics[width=2.5in]{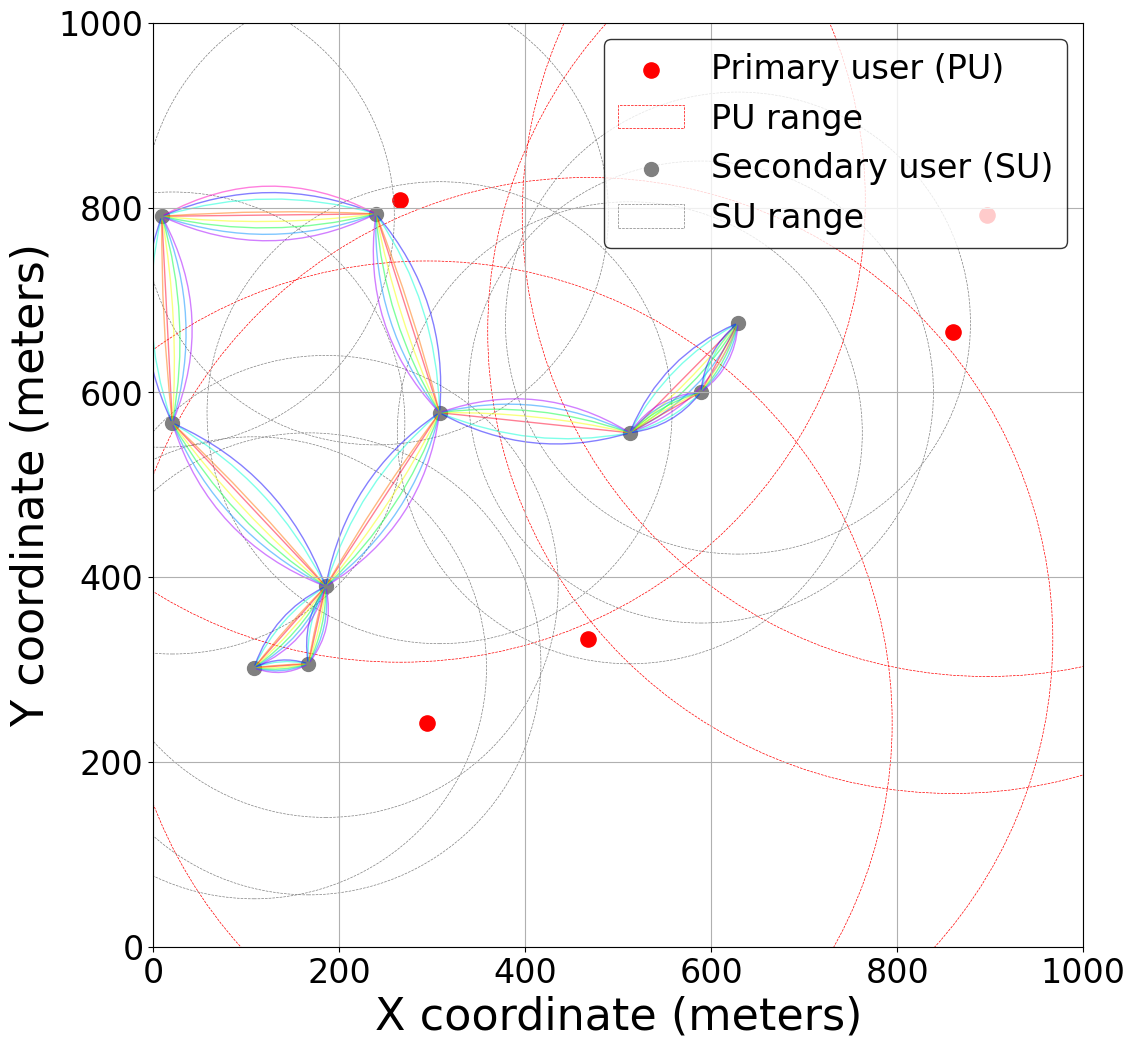}
	\caption{The multichannel topology discovery problem in a CRN.}
	\label{fig:mtd}
\end{figure}

In this paper, we extend the multichannel rendezvous framework to explicitly consider the underlying network topology, thus introducing the \emph{multichannel topology discovery problem}. We consider a CRN composed of primary and secondary users  distributed within the same geographical area, as illustrated in \rfig{mtd}. Primary users, depicted as red dots, occupy certain channels within their communication range (represented by red circles). Secondary users, depicted as grey dots, must avoid these occupied channels within their communication ranges (grey circles), defining their sets of available channels.

An edge exists between two secondary users if they lie within each other's communication range, leading to a network topology represented by a graph \(G=(V,E)\), with vertices corresponding to secondary users. Each secondary user has an available channel set determined by the absence of primary users in its vicinity. Edges between vertices reflect proximity and communication capability, with colored edges indicating shared channels available to connected users.

The \emph{multichannel topology discovery problem} thus requires secondary users to collaboratively and efficiently learn both the network graph's topology and the available channel sets of all secondary users through channel hopping over time. Addressing this problem enhances the ability of CRNs to self-organize and adapt dynamically to channel availability, improving overall communication performance and reliability.

The main contributions of this paper are summarized as follows:
\\

%\begin{enumerate}
%    \item We formulate the \emph{multichannel topology discovery problem}, extending the classical Multichannel Rendezvous Problem (MRP) to explicitly account for network topology and primary user interference.

%    \item We propose the \emph{pseudo-random sweep algorithm with forward replacement} for the multichannel topology discovery problem, which significantly reduces the expected time-to-discovery (ETTD) by mitigating the positive correlation between consecutive unsuccessful channel-hopping attempts.
%  We also provide a mathematical analysis that establishes a rigorous link between correlation in successive rendezvous trials and expected discovery time, thereby theoretically validating the advantage of pseudo-random channel hopping over sequential approaches.

\noindent (i)
%\item
We propose the \emph{pseudo-random sweep algorithm with forward replacement}, which significantly reduces the expected time-to-discovery (ETTD) by breaking the positive correlation between consecutive rendezvous attempts. A supporting mathematical analysis establishes a rigorous link between trial correlation and discovery time, theoretically validating the advantage of the pseudo-random sweep algorithm with forward replacement.
\\

\noindent (ii)  We introduce the \emph{threshold-based stick-together strategy}, which adaptively synchronizes user hopping sequences based on partial knowledge to accelerate topology discovery. We discuss how to choose the thresholds to reduce ETTD in the pseudo-random sweep algorithm with forward replacement.
\\

\noindent (iii)
%\item
We conduct comprehensive simulations to demonstrate that the proposed pseudo-random sweep algorithms significantly outperform the (sequential) sweep algorithms. Moreover, the ETTD achieved by the threshold-based stick-together strategy is even lower than that of the \({\bf \Pi}\)-algorithm \cite{CLC25}, the state-of-the-art channel-hopping algorithm for the MRP.

%\end{enumerate}

\bsec{Problem formulation}{problem}

Consider a CRN operating in discrete time, where time is divided into slots indexed by \(t=1,2,\ldots\). There are \(N\) globally labeled channels, numbered from 1 to \(N\), and \(K\) secondary users, numbered from 1 to \(K\). Because primary users may occupy some channels, each secondary user (hereafter simply referred to as a user) can only use a subset of the $N$ channels. We denote the set of available channels for user \(k\) as
\beq{avail0000}
{\bf c}_k = \{ c_{k,1}, \ldots, c_{k,n_k}\},
\eeq
where $n_k = |{\bf c}_k|$ is the number of available channels for user $k$.

The network topology is represented by a graph \(G=(V,E)\), where each vertex in \(V\) corresponds to a user. An edge \((i,j) \in E\) indicates that user \(i\) and user \(j\) are within each other’s communication range. The connectivity among the users is further described by the \(K\times K\) adjacency matrix \(A=(a_{i,j})\), where
$a_{i,j} =1$ if user $i$ and user $j$ are within communication range, and 0 otherwise.

The following assumptions are made:
\begin{description}
\item[(A1)] Connectivity: The graph $G$ is connected.
\item[(A2)] Common available channel: There is at least one channel
that is commonly available to the $K$ users, i.e.,
    \beq{avail1111}
\cap_{k=1}^K{\bf c}_k  \ne \varnothing.
\eeq

\item[(A3)] Clock synchronization: The clocks of the $K$ users are synchronized.
\item[(A4)] Communication assumption: A group of users can communicate using multihop at time \(t\) if they are tuned to the same channel and form a connected subgraph of \(G\). Once they have communicated, the topology of the connected subgraph is learned.
\end{description}

Initially, each user knows only its own available channel set. The {\em multichannel topology discovery problem} requires the \(K\) users to collaboratively learn both the adjacency matrix \(A\) and the available channel sets \(\mathbf{c}_k\) for all \(k = 1, 2, \ldots, K\).
%We note that the multichannel {\em topology} discovery problem is different from the multichannel {\em neighborhood} discovery problem \cite{chen2014heterogeneous,zeng2016robust}, which involves only the discovery of each user's neighbors through pairwise rendezvous.

In each time slot, every user hops to one of its available channels and communicates with others to exchange information. To facilitate topology discovery, each user maintains:
\begin{description}
\item[(i)] A {\em list of known users}, along with their available channel sets.
\item[(ii)] A {\em list of known edges} between the known users.
\end{description}
At the start, each user’s known user list contains only itself and its own available channel set, while its known edge list is empty.

At time \(t\), suppose a group of users hops onto the same channel \(c\) and forms a connected subgraph \(G_c\). By assumption (A4), these users can exchange information about what they have learned so far. After this exchange:
\begin{description}
\item[(i)] Each user in \(G_c\) updates its {\em known user list} to be the {\em union} of the known user lists of all users in \(G_c\).
\item[(ii)] Each user in \(G_c\) updates its {\em known edge list} to be the {\em union} of the known edge lists of all users in \(G_c\), plus the edges in \(G_c\) itself.
\end{description}
Note that it is possible that the group of users hopping onto the same channel \(c\) may form more than one connected subgraph. In that case, each connected subgraph can still update its knowledge as described above.
The {\em time-to-discovery (TTD)} is defined as the number of time slots required for all users to fully reconstruct the topology and learn all available channel sets. The goal is to design channel hopping algorithms that minimize TTD.

In this paper, we evaluate two key performance metrics:
\begin{description}
\item[(i)]  {\em Expected time-to-discovery (ETTD):} The average number of time slots required for complete topology discovery.
    \item[(ii)] {\em Maximum time-to-discovery (MTTD):} The worst-case number of time slots required for complete topology discovery.
\end{description}

\iffalse
One n\"aive approach for this problem is to ask each user randomly hop on one of its available channels at each time slot.
 adds the available channel sets
Such an approach is called the {\em random} algorithm.
\fi

\bsec{The baseline algorithms}{algorithms}

In this section, we review two known algorithms in the literature: the sweep algorithm in \rsec{sweep}
and the state-of-the-art ${\bf \Pi}$-algorithm in \rsec{consistent}.
These known algorithms serve as baseline algorithms for comparison in our experiments.

\bsubsec{The sweep algorithm}{sweep}

A simple approach for the multichannel topology discovery problem is to ask each user to hop on channel $t$ at time $t$ if channel $t$ is in its available channel set. If not, it remains idle (i.e., in sleep mode). Clearly, under the assumption in (A2),
there is a channel, say channel $t^*$, that is commonly available to every user. Since the clocks of the $K$ users are synchronized as described in (A3),  every user hops on channel $t^*$ at time $t^*$. From (A1) and (A4), all the $K$ users can communicate with each other (using multihop) at time $t^*$ to learn the adjacency matrix \(A\) and the available channel sets for all \(k = 1, 2, \ldots, K\). This simple approach guarantees that MTTD is bounded above $N$. We refer to this approach as the (sequential) sweep algorithm (cf. \cite{SynMAC}).

Based on the sweep algorithm, there exists at least one time slot during which all users hop on the same channel. Following the multichannel rendezvous literature (see, e.g., \cite{Book,GAP2019,Wang2022}), we define the time-to-rendezvous (TTR) as the number of time slots required for all users to hop on the same channel simultaneously. From (A4), we have
\beq{inequal1111}
TTD \le TTR.
\eeq
%Thus, channel hopping algorithms that minimize TTR are expected to perform well in reducing TTD.

In the basic sweep algorithm, a user remains idle at time \(t\) if channel \(t\) is not in its available channel set. To further reduce TTD, we consider two enhanced variants:

\begin{description}
\item[(i)] Sweep with random replacement: A user that would otherwise be idle at time \(t\) instead hops on a randomly selected channel from its available channel set.
\item[(ii)] Sweep with forward replacement: A user that would otherwise be idle at time \(t\)  instead hops on the available channel with the smallest index greater than \(t\), wrapping around modulo \(N\) if necessary.
\end{description}

To be precise, we outline the details of the sweep algorithm with random replacement in Algorithm
\ref{alg:sweepr} and the sweep algorithm with forward replacement in Algorithm
\ref{alg:sweepf}, respectively.

\begin{algorithm}
\caption{The sweep algorithm with random replacement}
\label{alg:sweepr}
\noindent {\bf Input}: A set of available channels \({\bf c} = \{c_1, \ldots, c_n\} \subseteq \{1,2, \ldots, N\}\), and a  pseudo-random permutation \(\pi\) of the set $\{1,2, \ldots, N\}$.

\noindent {\bf Output}: A channel hopping sequence \(\{c(t), t=1, 2, \ldots, N\}\) with \(c(t) \in {\bf c}\).

\noindent 1: If $t$ is in ${\bf c}$, set $c(t)=t$.

\noindent 2: Otherwise, randomly choose a channel $c$ from ${\bf c}$ and set $c(t)=c$.

\end{algorithm}

\begin{algorithm}
\caption{The sweep algorithm with forward replacement}
\label{alg:sweepf}
\noindent {\bf Input}: A set of available channels \({\bf c} = \{c_1, \ldots, c_n\} \subseteq \{1, 2, \ldots, N\}\), and a  pseudo-random permutation \(\pi\) of the set $\{1,2, \ldots, N\}$.

\noindent {\bf Output}: A channel hopping sequence \(\{c(t), t=1, 2, \ldots, N\}\) with \(c(t) \in {\bf c}\).

\noindent 1: If $t$ is in ${\bf c}$, set $c(t)=t$.

\noindent 2: Otherwise, set $c(t)=c_{i^*}$, where
\beq{sweepf1111} i^*={\rm argmin}_{1 \le i \le n}((c_i-t)\;\mod\;N).
\eeq

%\noindent 1: Generate CH sequence $\{c(t), t =1,2, \ldots, N\}$ with
%$c(t)=c_{i^*}$, where $i^*={\rm argmin}_{1 \le i \le n}((c_i-t)\;\mod\;N)$.

\end{algorithm}

\bsubsec{The ${\bf \Pi}$-algorithm}{consistent}

In this section, we briefly review the \({\bf \Pi}\)-algorithm from \cite{CLC25}.
It represents the state-of-the-art approach for minimizing the ETTR in the MRP when the clocks of the two users are synchronized.

\bdefin{pialg}({\em The ${\bf \Pi}$-algorithm})
For an available channel set \({\bf c} \subseteq \{1, 2, \ldots, N\}\), let \(\phi^{\min}({\bf c})\) denote the channel in \({\bf c}\) with the smallest index. Given a permutation \(\pi\) of \(\{1, 2, \ldots, N\}\), define \(\pi({\bf c})\) as the set of channels obtained by relabeling the elements of \({\bf c}\) according to \(\pi\). Let \({\bf \Pi} = \{\pi_1, \pi_2, \ldots\}\) be a sequence of such permutations. At time \(t\), the \({\bf \Pi}\)-algorithm selects the channel
\[
c(t) = \left( \pi_t^{-1} \circ \phi^{\min} \circ \pi_t \right)({\bf c}), \quad t = 1, 2, \ldots.
\]
If the sequence \({\bf \Pi}\) consists of independent, uniformly random permutations over all \(N!\) possibilities, the algorithm is referred to as the \emph{randomized \({\bf \Pi}\)-algorithm}.
\edefin

An important result for this randomized \({\bf \Pi}\)-algorithm (see Theorem 14 in \cite{CLC25}) is that the expected time-to-rendezvous (ETTR) for the MRP with two users is the inverse of the Jaccard index of their available channel sets, i.e.,
\[
\text{ETTR} = \frac{1}{J} = \frac{|{\bf c}_1 \cup {\bf c}_2|}{|{\bf c}_1 \cap {\bf c}_2|},
\]
where \({\bf c}_1\) and \({\bf c}_2\) are the available channel sets of the two users.
In view of this result, we expect that the randomized \({\bf \Pi}\)-algorithm also performs very well for the multichannel topology discovery problem. However, its implementation complexity is high, as it requires generating a sequence of pseudo-random permutations.

\begin{algorithm}
\caption{The randomized \({\bf \Pi}\)-algorithm}
\label{alg:pirandom}
\noindent {\bf Input}: A set of available channels \({\bf c} = \{c_1, \ldots, c_n\} \subseteq \{1, \ldots, N\}\), and a sequence of pseudo-random permutations \(\{\pi_1, \pi_2, \ldots\}\).

\noindent {\bf Output}: A channel hopping sequence \(\{c(t), t=1, 2, \ldots\}\) with \(c(t) \in {\bf c}\).

\noindent 1: \(c(t) = \big( \pi_t^{-1} \circ \phi_{\min} \circ \pi_t \big)({\bf c})\).
\end{algorithm}

\bsec{The Proposed algorithms}{prandom}

\bsubsec{The pseudo-random sweep algorithm with forward replacement}{prandomf}

In this section, we propose the pseudo-random sweep algorithm with forward replacement.
 All users share a common pseudo-random permutation $\pi$ of the set $\{1,2, \ldots, N\}$.
 At time $t$, a user attempts to hop on channel $\pi(t)$ if it is available.
If \(\pi(t)\) is unavailable, the user selects the next available channel with the smallest index greater than
$\pi(t)$, wrapping around modulo \(N\) if needed. The detailed procedure of this algorithm is presented in Algorithm~\ref{alg:psweep}.

\begin{algorithm}
\caption{The pseudo-random sweep algorithm with forward replacement}
\label{alg:psweep}
\noindent {\bf Input}: A set of available channels \({\bf c} = \{c_1, \ldots, c_n\} \subseteq \{1, 2,\ldots, N\}\), and a  pseudo-random permutation \(\pi\) of the set $\{1,2, \ldots, N\}$.

\noindent {\bf Output}: A channel hopping sequence \(\{c(t), t=1, 2, \ldots, N\}\) with \(c(t) \in {\bf c}\).

\noindent 1: ({\bf Pseudo-random sweep}) If $\pi(t)$ is in ${\bf c}$, set $c(t)=\pi(t)$.

\noindent 2: ({\bf Forward replacement}) Otherwise, set $c(t)=c_{i^*}$, where
\beq{psweepf1111} i^*={\rm argmin}_{1 \le i \le n}((c_i-\pi(t))\;\mod\;N).
\eeq
\end{algorithm}

The only difference between the sweep algorithm with forward replacement in \rsec{sweep} and the pseudo-random sweep algorithm in Algorithm~\ref{alg:psweep} is that the latter performs the sweep in a pseudo-random order. As such, the MTTD of Algorithm~\ref{alg:psweep} is bounded by \(N\), but its ETTD is significantly lower than that of the sweep algorithm with forward replacement. Simulation results in \rsec{sim} will show that its ETTD is comparable to that of the randomized ${\bf \Pi}$-algorithm. Moreover, since Algorithm~\ref{alg:psweep} requires only a single pseudo-random permutation, its implementation complexity is substantially lower than that of
the randomized ${\bf \Pi}$-algorithm.

\begin{figure}[ht]
\centering
\includegraphics[width=0.3\textwidth]{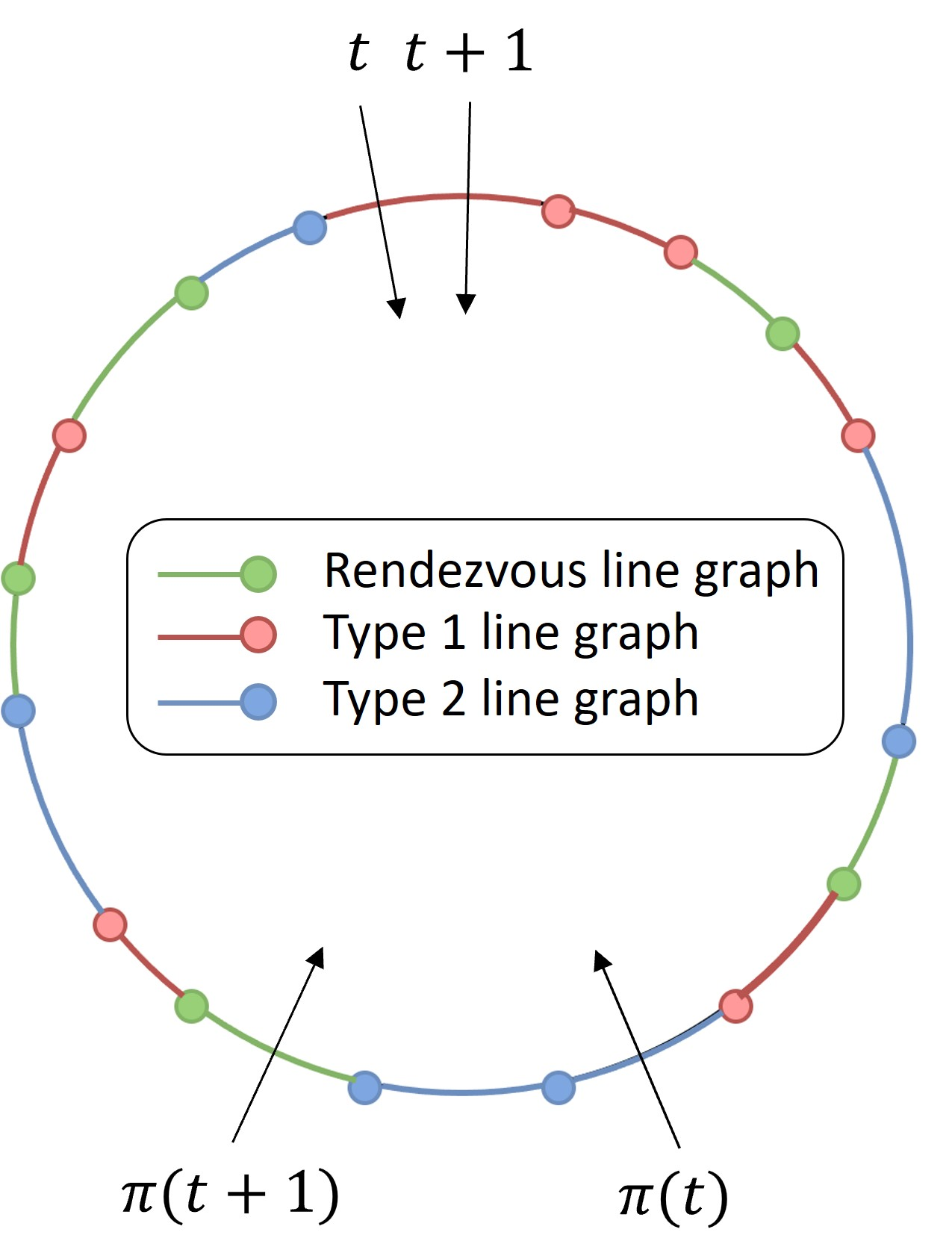}
\caption{A motivating illustration for using pseudo-random sweep.}
\label{fig:motivating}
\end{figure}

To provide insight into why the pseudo-random sweep algorithm is better than the (sequential) sweep algorithm,
let us consider the special case for $K=2$. Then the multichannel topology discovery problem is reduced to the
classical multichannel rendezvous problem with two users. Recall that ${\bf c}_1$ (resp.  ${\bf c}_2$) is the available channel set of user 1 (resp. user 2). Also, $n_1=|{\bf c}_1|$ (resp. $n_2=|{\bf c}_2|$) is the number of available channels of user 1 (resp. user 2). Let $n_{1,2}=|{\bf c}_1 \cap {\bf c}_2|$ be the number of common channels. As discussed in \cite{LSH}, we plot the available channels of the two users on a ring with $N$ nodes, as shown in \rfig{motivating}, where the green nodes are the common channels i.e., ${\bf c}_1 \cap {\bf c}_2$, the red nodes are the channels in user 1 that are not common channels, i.e.,  ${\bf c}_1 \backslash {\bf c}_2$, and
 the blue  nodes are the channels in user 2 that are not common channels, i.e.,  ${\bf c}_2 \backslash {\bf c}_1$.
These colored nodes partition the ring into $n_1+n_2-n_{1,2}$ line graphs.
We can classify the $n_1+n_2-n_{1,2}$ line graphs into three types: (i) a rendezvous line graph (marked in green in \rfig{motivating}): the forward end node of the line graph  is a common channel, (ii) a type 1 line graph (marked in red in \rfig{motivating}): the forward end node of the line graph is an available channel of user 1 that is not a common channel, and (iii) a type 2 line graph (marked in blue in \rfig{motivating}): the forward end node of the line graph is  an available channel of user 2 that is not a common channel.
Since there are $n_{1,2}$ common channels,
there are $n_{1,2}$ {\em rendezvous line graphs}. When a channel-hopping attempt
is in a rendezvous line graph, these two users rendezvous.
Otherwise, they do not rendezvous. If we view each channel-hopping attempt as a Bernoulli trial,
then the probability that a channel-hopping attempt results in a successful rendezvous in the (sequential) sweep algorithm
is the same as that in the pseudo-random sweep algorithm.
However, as illustrated in \rfig{motivating}, a failed Bernoulli trial in the (sequential) sweep algorithm is more likely to be followed by another failed trial, since channel-hopping attempts in time slots \(t\) and \(t+1\) are likely to fall within the same line graph. This implies that the Bernoulli trials in the (sequential) sweep algorithm are positively correlated. In contrast, the Bernoulli trials in the pseudo-random sweep algorithm are nearly independent, especially when \(N\) is large.

We support this insight with a mathematical analysis, presented in the following theorem.

\bthe{twostate}
%Consider a sequence of not necessarily independent Bernoulli random variables $\{X_1, X_2, \ldots\}$ with the same parameter $p$, i.e., $\pr( X_t=1)=p$ and $\pr( X_t=0)=1-p$ for all $t$.
Let \(\{X_t\}_{t\ge1}\) be a sequence of Bernoulli random variables generated by a stationary, homogeneous two-state Markov chain on \(\{0,1\}\) with
\[
\Pr(X_t=1)=p,\quad \Pr(X_t=0)=1-p,
\]
and transition probabilities
\[
p_{i,j} \;=\;\Pr\bigl(X_{t+1}=j \mid X_t=i\bigr),
\quad i,j\in\{0,1\}.
\]
Let $T=\inf\{t \ge 1: X_t=1\}$ be the first time that a successful trial occurs.
Also let
\[
\omega \;=\;\mathrm{Corr}(X_t,X_{t+1})
\]
be the Pearson correlation coefficient between consecutive trials.  Then
for fixed \(p\), $\ex[T]$ is a strictly increasing function of \(\omega\).
\ethe

%The proof of \rthe{twostate} is omitted here due to page limitations and can be found in our arXiv paper with the same title.

\bproof

We first derive the tail distribution $\pr(T>t)$ via the Markov property.
   By definition,
   \[
   \pr(T>t)
   =\pr\bigl(X_1=0,\,X_2=0,\,\dots,\,X_t=0\bigr).
   \]
   Stationarity gives \(\pr(X_1=0)=1-p\).  Then the homogeneous Markov property yields
\bear{psweepf3333}
   \pr(T>t)
   &=&\pr(X_1=0)\,\prod_{s=2}^t \pr\bigl(X_s=0\mid X_{s-1}=0\bigr) \nonumber\\
   &=&(1-p)\,p_{0,0}^{\,t-1}.
\eear

We then derive an expression for \(\omega\).
   By stationarity,
   \[
   \ex[X_t]=p,\quad {\rm Var}(X_t)=p(1-p).
   \]
   Also
  \bearn
   \ex[X_tX_{t+1}]
   &=&\pr(X_t=1,X_{t+1}=1)\\
   &=&\pr(X_t=1)\,\pr(X_{t+1}=1\mid X_t=1)\\
  & =&p\,p_{1,1}.
  \eearn
   Hence
   \[
   \omega
   =\frac{\ex[X_tX_{t+1}]-\ex[X_t]\ex[X_{t+1}]}{{\rm Var}(X_t)}
   =\frac{p\,p_{1,1}-p^2}{p(1-p)}
   =\frac{p_{1,1}-p}{1-p}.
   \]
   Meanwhile, the {\em detailed-balance} (stationary) equation for the two-state Markov chain is
   \[
   p\,(1-p_{1,1})
   \;=\;(1-p)\,(1-p_{0,0}).
   \]
   This implies that
  \[
   p_{1,1}
   =1-\frac{1-p}{p}\,(1-p_{0,0}).
   \]
   Substituting into the expression for \(\omega\) gives
  \bear{psweepf5555}
   \omega
   &=&\frac{1-\tfrac{1-p}{p}(1-p_{0,0})-p}{1-p} \nonumber\\
   &=&\frac{p_{0,0}-(1-p)}{p} \nonumber\\
   &=&\frac{p_{0,0}}{p}-\frac{1-p}{p}.
\eear

   From \req{psweepf3333}, \(\pr(T>t)\) is a strictly increasing function of \(p_{0,0}\).  From \req{psweepf5555}, \(\omega\) is an affine increasing function of \(p_{0,0}\) (for fixed \(p\)).  Therefore, for any fixed \(p\),
   $\Pr(T>t)$ is a strictly increasing function of \(\omega\).
From the tail sum formula for the expectation, i.e.,
   $$\ex[T]=\sum_{t=0}^\infty \pr (T >t),$$
  we conclude that $\ex[T]$ is a strictly increasing function of \(\omega\).
   \eproof

\bsubsec{The stick-together strategy}{stick}

In the pseudo-random sweep algorithm with forward replacement (Algorithm \ref{alg:psweep}), users follow a fixed channel hopping sequence, regardless of any learned information about other users or their available channel sets.
However, inspired by the “stick-together” strategy introduced in \cite{ToN2017}, we consider a variant in which users adapt their channel hopping sequences based on known information.

Specifically, suppose that at time \(t\), a user knows a subset \(S\) of the \(K\) users and their corresponding available channel sets. The user can then compute the {\em stick-together channel set} \({\bf c}_{\rm st}=\cap_{k \in S} {\bf c}_k\), and use this as the basis for its channel hopping sequence from time \(t\) onward.

\iffalse
This variant is described in Algorithm~\ref{alg:psweepst}.

\begin{algorithm}
\caption{The pseudo-random sweep algorithm with the stick-together strategy}
\label{alg:psweepst}
\noindent {\bf Input}: A list of known users along with their available channel sets, and a  pseudo-random permutation \(\pi\).

\noindent {\bf Output}: A channel hopping sequence \(\{c(t), t=1, 2, \ldots\}\) with \(c(t) \in {\bf c}\).

\noindent 1: At time $t$, the user computes the stick-together channel set \({\bf c}_{\rm st}\) as the intersection of the available channel sets of the known users.

\noindent 2: Use the pseudo-random sweep algorithm with forward replacement (Algorithm \ref{alg:psweep}) and
the available channel set \({\bf c}_{\rm st}\) to generate $c(t)$.
\end{algorithm}
\fi

\begin{figure}[ht]
\centering
\includegraphics[width=0.15\textwidth]{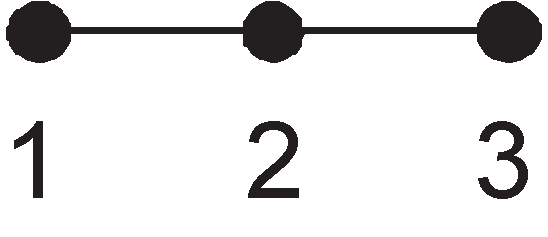}
\caption{A line graph with three nodes.}
\label{fig:linegraph}
\end{figure}

To illustrate the benefit of the stick-together strategy,  consider the line graph topology illustrated in \rfig{linegraph}, which consists of three users (nodes) arranged in a linear fashion: user 1 is connected to user 2, and user 2 is connected to user 3. Suppose that user 1 and user 2 successfully rendezvous at some point in time and subsequently adopt the stick-together strategy. That is, they update their channel hopping sequences to use a common set of available channels—the intersection of their individual channel sets.

Now, when user 2 later rendezvous with user 3, the discovery of this additional connection allows user 2 to infer the existence of both links in the line graph (1-2 and 2-3). Because user 1 is already synchronized with user 2 through the stick-together channel set, user 2 can share the newly discovered information with user 1, thereby enabling all three users to reconstruct the entire line topology.

In contrast, if users 1 and 2 do not adopt the stick-together strategy after their rendezvous, they will continue to use their original, possibly disjoint hopping sequences. As a result, even after user 2 rendezvous with user 3, user 1 has no coordinated mechanism for learning about user 3—thus, the full topology remains undiscovered from user 1’s perspective.

This example illustrates how the stick-together strategy can enhance both the efficiency and completeness of topology discovery by promoting information sharing and coordination across the network.

However, there is a tradeoff. If the number of common channels between user 1 and user 2 is significantly smaller than the number of available channels for user 2, then restricting user 2 to the smaller stick-together channel set may reduce its probability of rendezvousing with user 3. In such cases, the stick-together strategy could actually increase the TTD by limiting exploration flexibility. Therefore, while the stick-together strategy can accelerate network-wide discovery in some settings, it may be counterproductive when adopted too early or among users with limited overlap in available channels.

Based on the above argument,  using the stick-together strategy in Algorithm \ref{alg:psweep} is expected to perform well when the number of common channels is large, but its performance degrades significantly when the number of common channels is small. To address this limitation, we propose a {\em threshold-based stick-together strategy}, in which the stick-together strategy is used  only when the size of \({\bf c}_{\rm st}\) is at least a predefined threshold \(n_{\rm TH}\) and the number of known users is at least a predefined threshold \(k_{\rm TH}\). The detailed procedure is described in Algorithm \ref{alg:psweepstth}.

\begin{algorithm}
\caption{Pseudo-random sweep algorithm with threshold-based stick-together strategy}
\label{alg:psweepstth}
\noindent {\bf Input}: A set of available channels \({\bf c} = \{c_1, \ldots, c_n\}\) of the user,  a list of known users along with their available channel sets, a pseudo-random permutation \(\pi\), and two thresholds \(n_{\rm TH}\) and \(k_{\rm TH}\).

\noindent {\bf Output}: A channel hopping sequence \(\{c(t), t=1, 2, \ldots\}\) with \(c(t) \in {\bf c}\).

\begin{enumerate}
%\item Initially, the user knows only its own available channel set \({\bf c}\), and sets the stick-together channel set \({\bf c}_{\rm st} = {\bf c}\).

\item At time \(t\), let $S$ be the set of known users.  The user computes the  stick-together channel set \({\bf c}_{\rm st}\) as the intersection of the available channel sets of all known users.

\item If \(|{\bf c}_{\rm st}| \ge n_{\rm TH}\) and  $|S| \ge k_{\rm TH}$, then
use
the pseudo-random sweep algorithm with forward replacement in
Algorithm~\ref{alg:psweep} with the input \({\bf c}_{\rm st}\) to determine \(c(t)\).

\item Otherwise, use
the pseudo-random sweep algorithm with forward replacement in
Algorithm~\ref{alg:psweep} with the input \({\bf c}\) to determine \(c(t)\).
\end{enumerate}
\end{algorithm}

\bsec{Simulations}{sim}

In all our experiments, we set the number of channels to \( N = 256 \) and the number of secondary users (SUs) to \( K = 100 \). To generate the network topology, we randomly distribute these 100 SUs within a square of dimensions \( 1,000 \) m \( \times \) \( 1,000 \) m. The communication range between two SUs is set to \( 250 \) meters. Consequently, an edge is established between two SUs if the distance between them is within the communication range.
For our experiments, we generate 1,000 topologies (represented as adjacency matrices) that satisfy the connectivity assumption stated in (A1).

To ensure that the nonempty intersection assumption in (A2) holds, we randomly select a subset of common channels, denoted as \( {\bf c}_{\rm com} \). The set of channels allocated to primary users (PUs) is then defined as
\[
{\bf c}_{\rm PU} = \{1,\ldots, N\} \backslash {\bf c}_{\rm com}.
\]
The number of PUs is set to 50, and these PUs are randomly distributed within the same square area as the SUs. The communication range of each PU is set to twice that of an SU, i.e., \( 500 \) meters. If a PU has no SUs within its communication range, it does not interfere with SU communication and is therefore removed from the set of PUs.

Let \( P \) be the number of remaining PUs after this filtering process. The channels in \( {\bf c}_{\rm PU} \) are then assigned to these \( P \) PUs as evenly as possible. A PU is considered an {\em interfering PU} to an SU if the SU falls within the PU's communication range.
The available channel set for an SU consists of the remaining channels from the total \( N \) channels that are not occupied by its interfering PUs. Given this channel assignment strategy, the set of common channels available to all \( K \) SUs is precisely \( {\bf c}_{\rm com} \).

The ETTD is calculated as the average over all 1,000 topologies. To measure the MTTD, we divide the 1,000 topologies into 100 batches, each containing 10 topologies. Let $MTTD_i$ denote the MTTD measured in
the $i^{th}$ batch, where $i=1,2, \ldots, 100$. The final MTTD value is then obtained by averaging $MTTD_i$ over the 100 batches.

\begin{figure}[!t]
	\centering
	\includegraphics[width=2.5in]{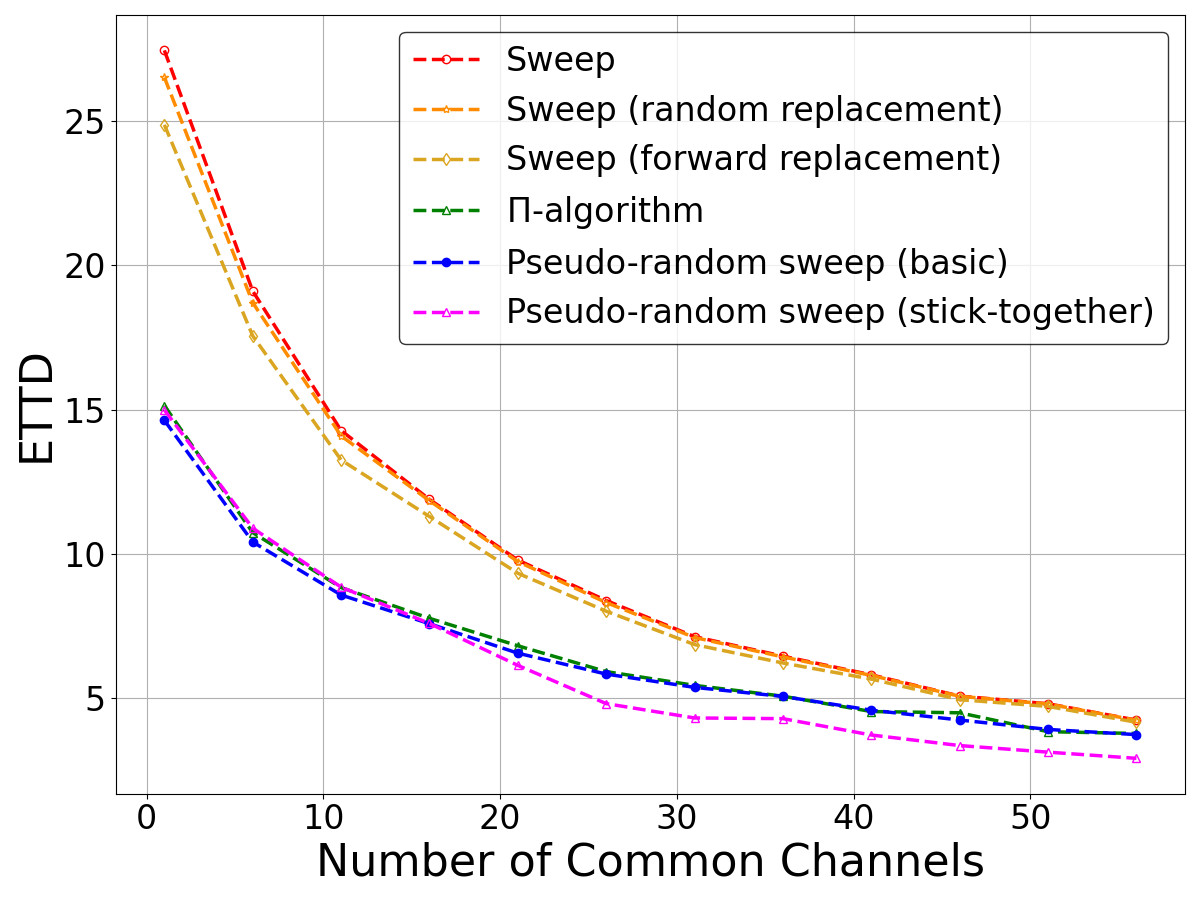}
	\caption{The ETTD comparison of the six algorithms.}
	\label{fig:ettd_compare}
\end{figure}

In \rfig{ettd_compare}, we first compare the ETTD of the sweep algorithm (marked as Sweep), the sweep algorithm with random replacement
(marked as Sweep (random replacement)), the sweep algorithm with forward replacement
(marked as Sweep (forward replacement)), the randomized ${\bf \Pi}$-algorithm (marked as ${\bf \Pi}$-algorithm),
the pseudo-random sweep algorithm with forward replacement in Algorithm \ref{alg:psweep} (marked as Pseudo-random sweep (basic)),
and
the pseudo-random sweep algorithm with the threshold-based stick-together strategy in Algorithm \ref{alg:psweepstth} (marked as
Pseudo-random sweep (stick-together)). The thresholds used in Algorithm \ref{alg:psweepstth} are $n_{\rm TH}=5$ and $k_{\rm TH}=30$.
%We consider three cases for Algorithm \ref{alg:psweepstth}:
%(i) $n_{\rm TH}=\infty$ and $k_{\rm TH}=\infty$, (ii) $n_{\rm TH}=0$ and $k_{\rm TH}=0$,  and (iii) $n_{\rm TH}=30$ and $k_{\rm TH}=30$. Case 1 corresponds to the pseudo-random sweep algorithm with forward replacement in Algorithm \ref{alg:psweep}, i.e., without using the stick-together strategy.  On the other hand, the  stick-together strategy is always used in Case 2.
As shown in the figure, the ETTDs of the three sweep-based algorithms are similar, indicating that the improvement from using channel replacement is relatively small. In contrast, the ETTD of the pseudo-random sweep algorithm with forward replacement is significantly lower than those of the three sweep algorithms. The rationale behind this, as explained in \rthe{twostate}, is that the Bernoulli trials of the (sequential) sweep algorithm are positively correlated while the Bernoulli trials of the pseudo-random sweep algorithm are nearly independent.
We also note that the ETTD of the pseudo-random sweep algorithm with forward replacement coincides with that of the randomized \({\bf \Pi}\)-algorithm, the state-of-the-art approach for the MRP.
The ETTD of the pseudo-random sweep algorithm with the threshold-based stick-together strategy is even lower than that of the randomized \({\bf \Pi}\)-algorithm.

\begin{figure}[!t]
	\centering
	\includegraphics[width=2.5in]{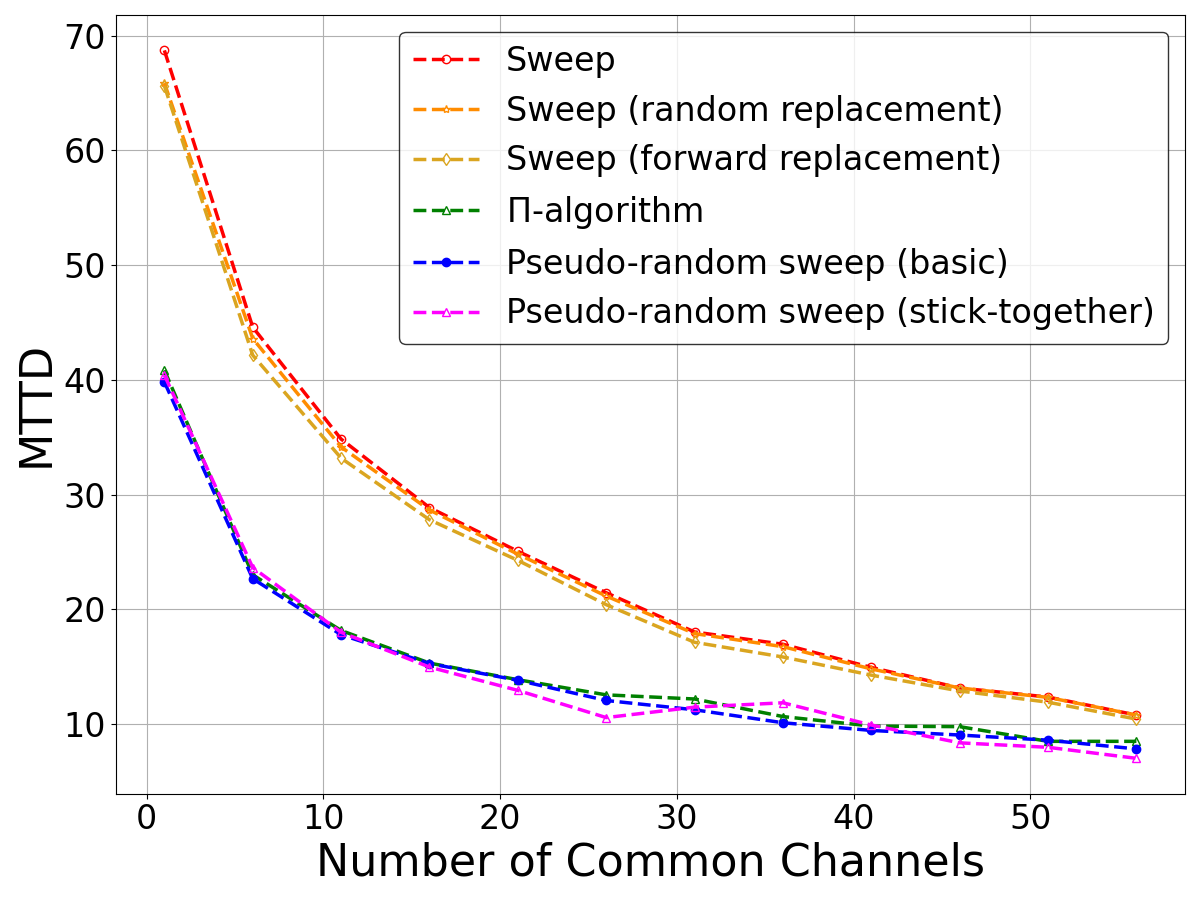}
	\caption{The MTTD comparison of the six algorithms.}
	\label{fig:mttd_compare}
\end{figure}

In \rfig{mttd_compare}, we compare the (measured) MTTD of these six algorithms. The results are similar to those in
\rfig{ettd_compare}. However, the gain of using the stick-together strategy is very small in this figure.

\bsec{Conclusion}{con}

In this paper, we introduced and studied the \emph{multichannel topology discovery problem}, an important generalization of the classical multichannel rendezvous problem that explicitly considers network topology and primary user interference. We proposed the \emph{pseudo-random sweep algorithm with forward replacement}, along with the  \emph{threshold-based stick-together strategy}, to enhance the efficiency of topology discovery. Our mathematical analysis established the direct relationship between the correlation of rendezvous trials and the expected discovery time, providing a rigorous justification for the effectiveness of pseudo-random methods. Extensive simulations demonstrated significant improvements in both expected and maximum time-to-discovery compared to the sweep algorithms.

Future work includes extending the pseudo-random sweep algorithm with forward replacement to the asynchronous setting, where user clocks are not synchronized. It appears plausible to adapt the dimension reduction technique from \cite{LSH} to this context. Another direction is to explore more realistic communication models beyond assumption (A4), potentially incorporating lower-layer protocol behaviors.

%\bibliographystyle{IEEEtran}
%\bibliography{MathMRP}

% Generated by IEEEtran.bst, version: 1.14 (2015/08/26)

\end{document}